\documentclass[doublecol]{epl2}
\usepackage{amssymb,amsmath,graphicx,setspace,color,rotating,subfigure,url,multirow}
\title{An improved HeatS+ProbS hybrid recommendation algorithm based on heterogeneous initial resource configurations}
\shorttitle{An improved HeatS+ProbS hybrid recommendation algorithm}

\author{Chuang Liu\inst{1,2,3} \and Wei-Xing Zhou\inst{1,2,4,5}\footnote{e-mail: wxzhou@ecust.edu.cn}}
\shortauthor{C. Liu \& W.-X. Zhou}

\institute{
  \inst{1} School of Business, East China University of Science and Technology, Shanghai 200237, China\\
  \inst{2} Engineering Research Center of Process Systems Engineering (Ministry of Education), East China University of Science and Technology,   Shanghai 200237, China\\
  \inst{3} Department of Physics, University of Fribourg - Chemin du Muse 3, CH-1700 Fribourg, Switzerland\\
  \inst{4} Research Center for Econophysics, East China University of Science and Technology, Shanghai 200237, China\\
  \inst{5} School of Science, East China University of Science and Technology, Shanghai 200237, China
}

 \pacs{89.75.Hc}{Networks and genealogical trees}
 \pacs{87.23.Ge}{Dynamics of social systems}
 \pacs{89.65.Gh}{Economics; econophysics, financial markets, business and management}

\abstract{Network-based recommendation algorithms for user-object link predictions have achieved significant developments in recent years. For bipartite graphs, the reallocation of resource in such algorithms is analogous to heat spreading (HeatS) or probability spreading (ProbS) processes. The best algorithm to date is a hybrid of the HeatS and ProbS techniques with homogenous initial resource configurations, which fulfills simultaneously high accuracy and large diversity. We investigate the effect of heterogeneity in initial configurations on the HeatS+ProbS hybrid algorithm and find that both recommendation accuracy and diversity can be further improved in this new setting. Numerical experiments show that the improvement is robust.}

\begin{document}

\maketitle

\section{\label{S1:Intro}Introduction}

In recent years, the huge data sets available in natural, social and information sciences have witnessed the flourish of complex network analysis \cite{Albert-Barabasi-2002-RMP,Newman-2003-SIAMR,Boccaletti-Latora-Moreno-Chavez-Hwang-2006-PR}. In most cases, the data are recorded as snapshots. The underlying mechanisms of network evolution are usually unknown, which is true in most situations even when the growth dynamics of networks are recorded. Therefore, one has to predict missing links in incompletely recorded networks or future links, which has important scientific and practical significance \cite{Sarukkai-2000-CN,LibenNowell-Kleinberg-2007-JASIST,Clauset-Moore-Newman-2008-Nature}. To date, various methods have been proposed and developed for link prediction in different fields \cite{Zhu-2001-LNAI,Zhu-Hong-Hughes-2002-LNCS,Marchette-Priebe-2008-CSDA,Zhou-Lu-Zhang-2009-EPJB,Lu-Jin-Zhou-2009-PRE,Lu-Zhou-2010-EPL,Liu-Lu-2010-EPL}.

As a special case of complex networks, bipartite graphs are quite common especially in social sciences. In everyday life, people buy books, articles for daily uses, and foods from online or convenience stores, collect online movies and music, choose restaurants and resorts, invest stocks and derivatives, and so on \cite{Resnick-Varian-1997-CACM}. In medical science, scientists try to unveil the unknown interaction mechanisms between huge numbers of drugs and targets \cite{Yildirim-Goh-Cusick-Barabasi-Vidal-2007-NBt}, and predicting possible drug-target links is of crucial importance in drug design. It is often necessary to make choices without sufficient personal experience of the alternatives. Recommender systems are mainly aimed at providing link predictions for such systems.

There are many recommender systems designed for different systems. One of the most successful methods for recommender systems is based on the collaborative filtering technique \cite{Goldberg-Nichols-Oki-Terry-1992-CACM}, which has a large number of variants \cite{Herlocker-Konstan-Terveen-Riedl-2004-ACMtis} and their hybrids \cite{Burke-2002-UMUAI}. Recently, a lot of efforts in the physics community have been devoted to design recommendation algorithms on bipartite graphs \cite{Zhou-Ren-Medo-Zhang-2007-PRE,Jia-Liu-Sun-Wang-2008-PA,Zhou-Su-Liu-Jiang-Wang-Zhang-2009-NJP,Liu-Deng-2009-PA,Zhou-Kuscsik-Liu-Medo-Wakeling-Zhang-2010-PNAS}, where the hybrid algorithm combining the heat spreading (HeatS) and probability spreading (ProbS) algorithms is found to achieve simultaneously higher recommendation accuracy and greater diversity \cite{Zhou-Kuscsik-Liu-Medo-Wakeling-Zhang-2010-PNAS}. In this work, we propose an improved HeatS+ProbS algorithm by considering the heterogeneity in initial source configurations.

\section{Algorithms}

Generally, recommender systems are designed based on bipartite user-object graphs ${\cal{G}}({\bf{u}},{\bf{o}},E)$, which contain users ${\bf{u}}=\{u_1,u_2,\cdots,u_m\}$, objects ${\bf{o}}=\{o_1,o_2,\cdots,o_n\}$, and links  $E=\{e_{i\alpha}:u_i\in {\bf{u}}, o_\alpha\in {\bf{o}}\}$. A link is drawn between $u_i$ and $o_\alpha$ if user $u_i$ has collected object $o_\alpha$. For readability, we use $i,j$ for the subscripts of users and $\alpha,\beta$ for objects. The user-object bipartite graph can be presented by an $m \times n$ adjacent matrix $A$, where $a_{i\alpha}=1$ if user $u_i$ has collected object $o_\alpha$ and $a_{i\alpha}=0$ otherwise.

The resource reallocation process for each user in the network-based recommendation algorithms can be expressed using a single equation
\begin{equation}
 {\bf{f}} = W{\bf{f}}_0
 \label{Eq:NBR:fWf0}
\end{equation}
where ${\bf{f}}_0=[f_{1,0}^i, \cdots, f_{n,0}^i]$ is the initial configuration of resource on objects, $W$ is the resource reallocation matrix, and ${\bf{f}}=[f_{1}^i, \cdots, f_{n}^i]$ is the final configuration of resource on objects. The objects are sorted in a descending order and a certain number of objects with the highest final resources that have not been collected by user $u_i$ are recommended to him. After one knows the resource reallocation matrix $W$ and the initial configuration ${\bf{f}}_0$ on objects, the recommendation algorithm is determined.

In the heat spreading algorithm \cite{Zhou-Kuscsik-Liu-Medo-Wakeling-Zhang-2010-PNAS}, a less popular object with low degree will obtain larger final resource and the recommendation list is diverse, where the resource reallocation matrix is
\begin{equation}
 W_{\alpha\beta} = \frac{1}{k_\alpha}\sum_{i=1}^m \frac{a_{i\alpha}a_{i\beta}}{k_i},
 \label{Eq:HeatS:WH}
\end{equation}
where $k_\alpha$ is the degree of $o_\alpha$ and $k_i$ is the degree of $u_i$. In contrast, a popular object with high degree will have more final resource in the ProbS algorithm and the recommendation list is accurate, where the resource reallocation matrix is \cite{Zhou-Ren-Medo-Zhang-2007-PRE}
\begin{equation}
 W_{\alpha\beta} = \frac{1}{k_\beta}\sum_{i=1}^m \frac{a_{i\alpha}a_{i\beta}}{k_i}.
 \label{Eq:ProbS:WP}
\end{equation}
In order to solve the apparent accuracy-diversity dilemma of recommender systems, a hybrid algorithm has been proposed \cite{Zhou-Kuscsik-Liu-Medo-Wakeling-Zhang-2010-PNAS}, which combines these two algorithms as follows
\begin{equation}
 W_{\alpha\beta} = \frac{1}{k_\alpha^{1-\lambda}k_\beta^\lambda}\sum_{i=1}^m \frac{a_{i\alpha}a_{i\beta}}{k_i}.
 \label{Eq:H+P:W}
\end{equation}
The elegant hybrid algorithm results in higher accuracy and greater diversity when the parameter $\lambda$ is tuned to around an optimal value.

The initial resource vector $\mathbf{f}_0$ in many network based recommendation algorithms, including the HeatS+ProbS hybrid algorithm, is determined as follows \cite{Zhou-Ren-Medo-Zhang-2007-PRE,Jia-Liu-Sun-Wang-2008-PA,Zhou-Su-Liu-Jiang-Wang-Zhang-2009-NJP,Zhou-Kuscsik-Liu-Medo-Wakeling-Zhang-2010-PNAS}
\begin{equation}
 f_\alpha^i = a_{i\alpha}.
 \label{Eq:NBR:f0:aij}
\end{equation}
That is to say, if object $o_\alpha$ has been collected by $u_i$, then its initial resource is one, otherwise it is zero. It has been shown that a heterogeneous initial configuration of resource
\begin{equation}
 f_\alpha^i = a_{i\alpha}k_\alpha^\eta
 \label{Eq:NBR:f0:aij:ko}
\end{equation}
can improve the recommendation accuracy of the ProbS algorithm \cite{Zhou-Jiang-Su-Zhang-2008-EPL}. The aim of this Letter is to investigate the effect of the initial resource configuration on the recommendation performance (accuracy and diversity) of the HeatS+ProbS hybrid algorithm.

\section{Data}

Two benchmark datasets have been adopted to test the performance of the recommendation algorithm. The first dataset, MovieLens, is downloaded from the website of GroupLens Research \cite{Konstan-Miller-Maltz-Herlocker-Gordon-Riedl-1997-CACM}. MovieLens' users rank movies at five discrete levels from 1 to 5. It contains $n=1682$ movies (objects), $m=943$ users, and 100,000 ratings. If the rating of movie $o_\alpha$ made by user $u_i$ is no less than 3, we argue that $u_i$ collected $o_\alpha$. This results in 82520 user-object pairs and the sparsity of the bipartite network is 0.0582. The second dataset, Netflix, is a randomly selected subset of the huge dataset provided for the Netflix Prize \cite{Bennett-Lanning-2007-Netflix}. It consists of $n=6000$ objects, $m=10000$ users, and 701749 links after a coarse-graining map from the five-level rating to the unary form. The sparsity of the bipartite network is 0.0117.

In order to investigate the performance of the proposed recommendation algorithm, the links in each data set are randomly divided into two subsets. The training set contains 90\% links while the probe set $E_P$ contains the remaining 10\% links. The algorithm is implemented using the training set to make recommendations, which are compared with the links in the probe set for performance (accuracy and diversity) \cite{Zhou-Kuscsik-Liu-Medo-Wakeling-Zhang-2010-PNAS}.

\section{Accuracy of recommendation}

\begin{figure*}[htb]
  \centering
  \includegraphics[width=5.5cm]{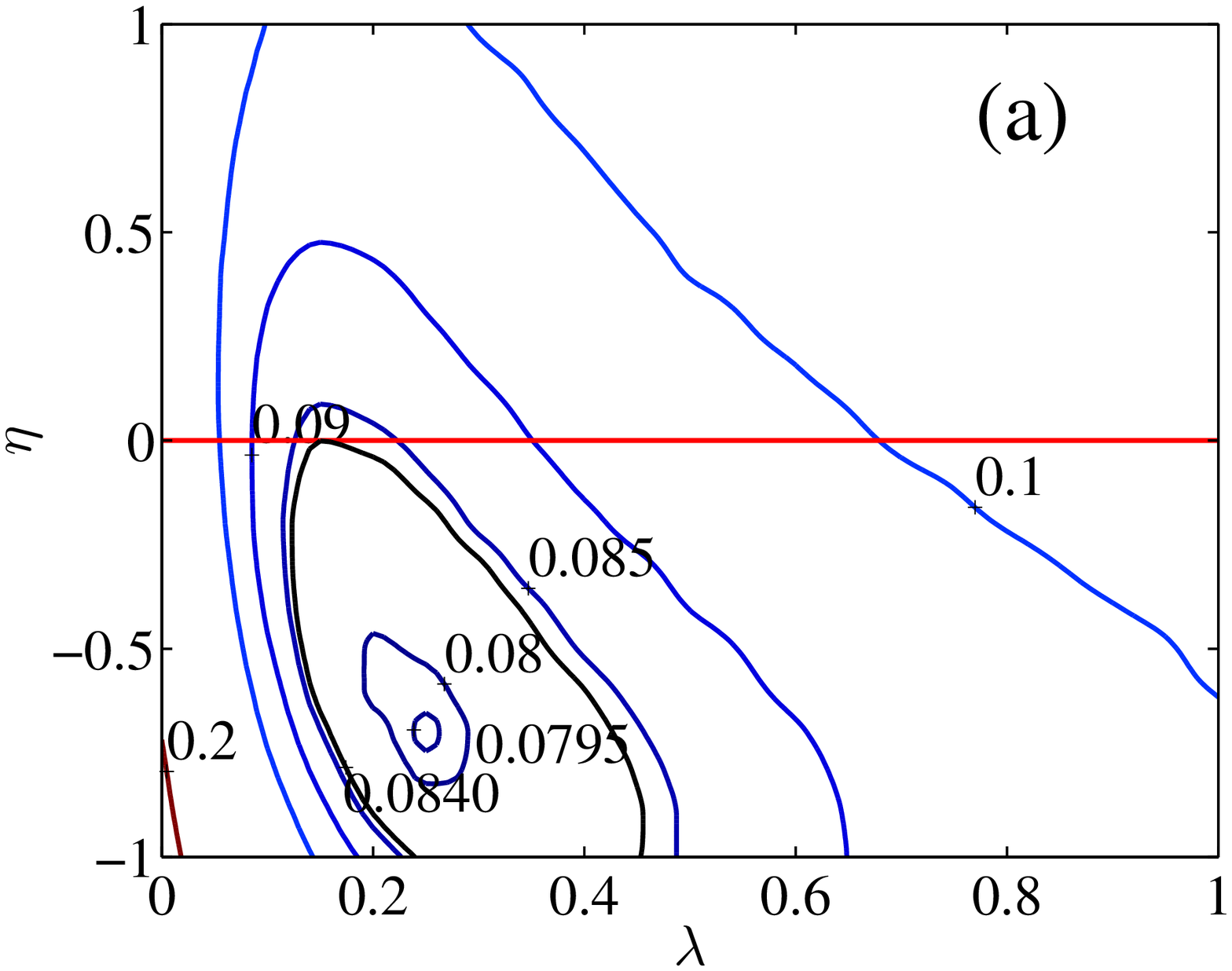}
  \includegraphics[width=5.5cm]{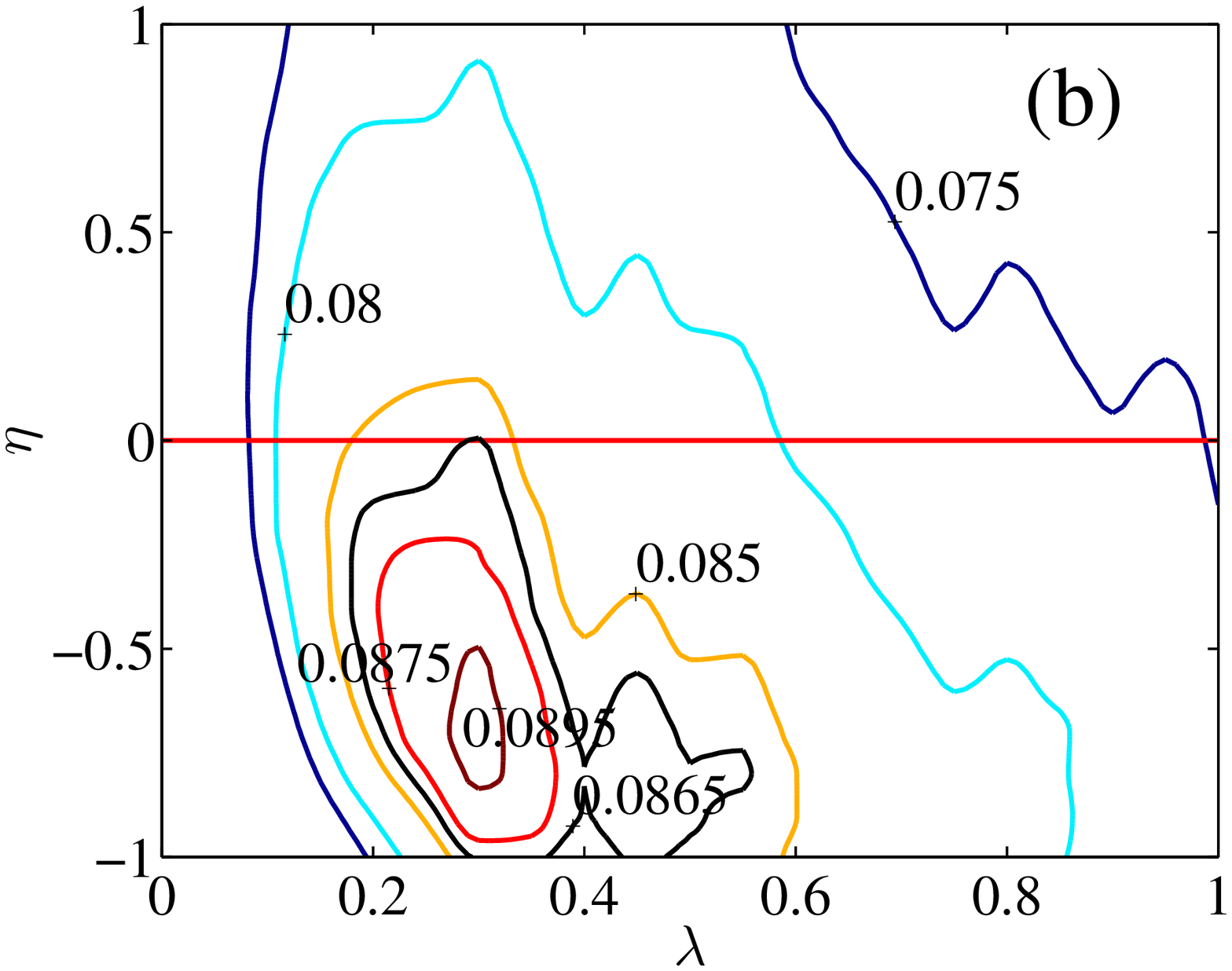}
  \includegraphics[width=5.5cm]{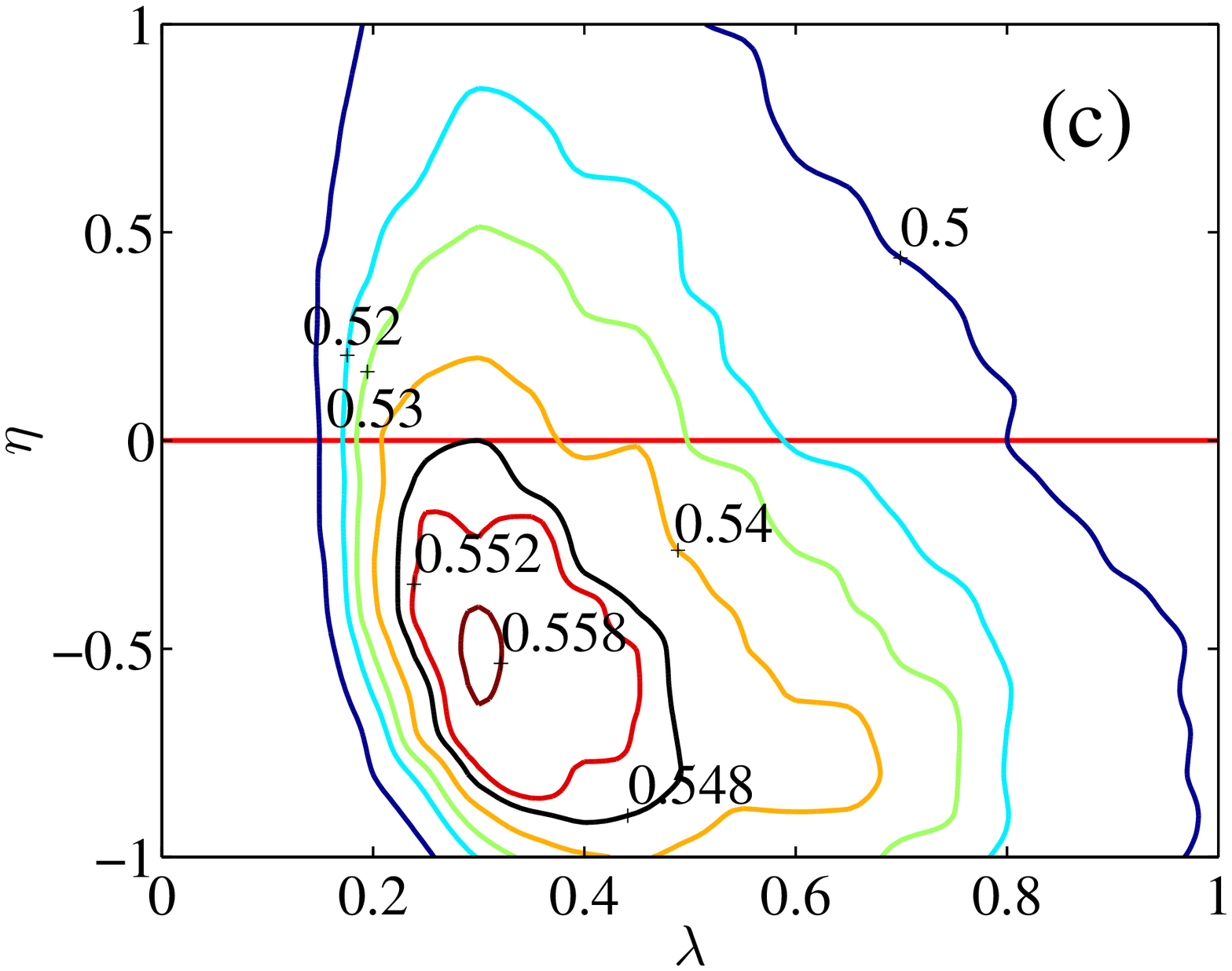}
  \includegraphics[width=5.5cm]{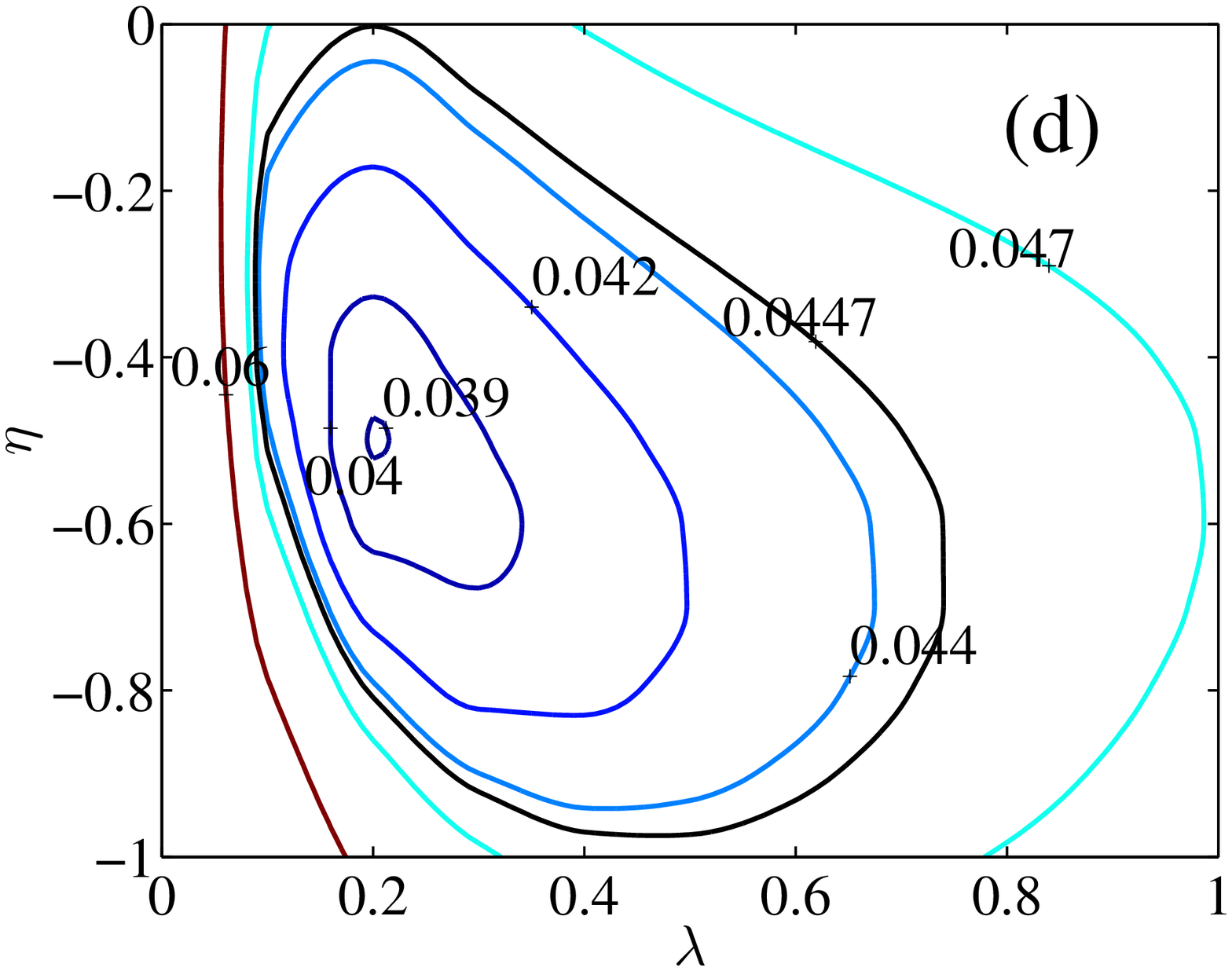}
  \includegraphics[width=5.5cm]{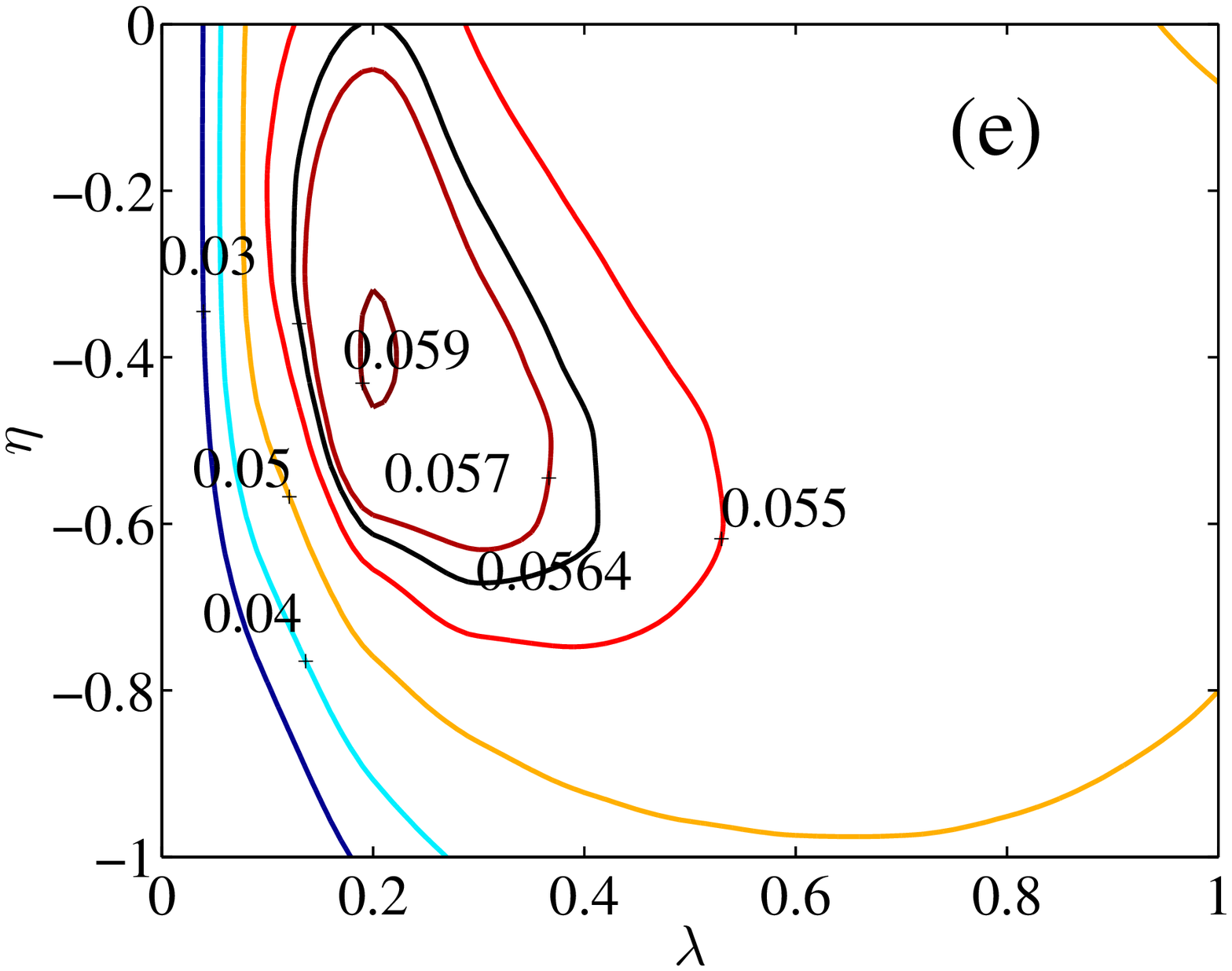}
  \includegraphics[width=5.5cm]{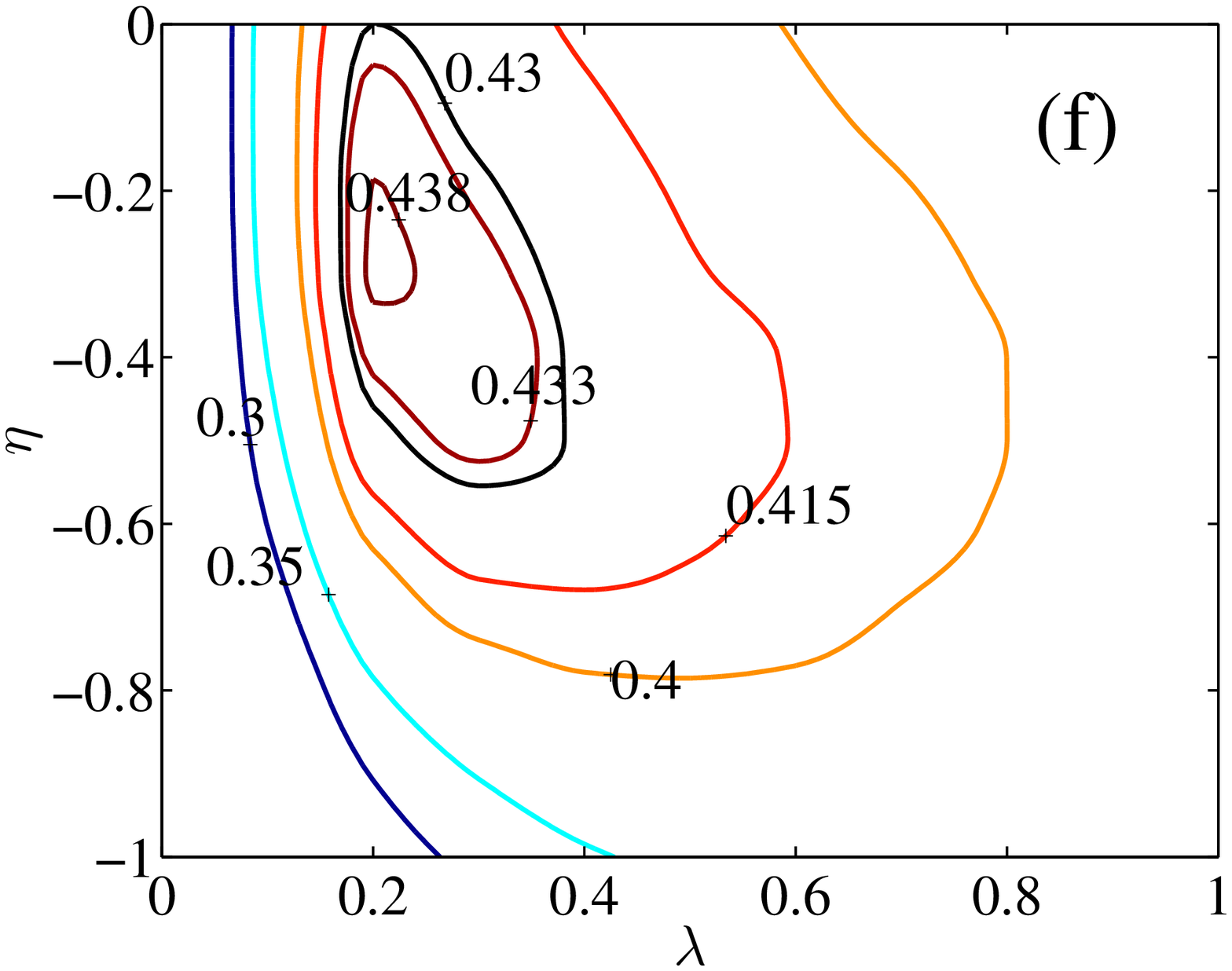}
  \caption{\label{Fig:HPIC:Accuracy} (Color online) Contour plots of the three recommendation accuracy measures with respect to the two parameters $\lambda$ and $\eta$. The first row is for the MovieLens data and the second row for the Netflix data. The three columns correspond to the ranking score $r(\lambda,\eta)$, the precision $P(\lambda,\eta)$, and the recall $R(\lambda,\eta)$. For the precision and recall, the length of the recommendation list is $L=50$. The contour lines tangent to the line $\eta=0$ are the HP lines for the three accuracy measures.}
\end{figure*}

We utilize three measures for the quantification of the recommendation accuracy. The first measure is the ranking score, which is defined as follows \cite{Zhou-Ren-Medo-Zhang-2007-PRE}
\begin{equation}
 r = \frac{1}{|E_P|}\sum_{i\alpha\in E_P}\frac{ q_{i\alpha}}{n-k_i}
 \label{Eq:rankingscore}
\end{equation}
where $|E_P|$ is the number of links in the probe set, and $q_{i\alpha}$ is the position that $o_\alpha$ placed in user $u_i$'s recommendation list. If the objects ranking from $q_1$ to $q_2$ in the list have the same score as $o_\alpha$, $q_{i\alpha}=(q_1+q_2)/2$ \cite{Zhou-Kuscsik-Liu-Medo-Wakeling-Zhang-2010-PNAS}. The smaller is $r$, the more accurate is the algorithm.

Plots (a) and (d) of fig.~\ref{Fig:HPIC:Accuracy} present the contours of the $r(\lambda,\eta)$ functions for the MovieLens data and the Netflix data, showing the dependence of the ranking score $r$ as a function of the two parameters $\lambda$ and $\eta$. Note that the results for the HeatS+ProbS hybrid algorithm are given by $\eta=0$. For simplicity, we call the contour line whose ranking score is the minimum obtained by the optimal $\lambda$ in the HeatS+ProbS hybrid algorithm as the {\em{HP line}} for ranking score,
\begin{equation}
 r(\lambda,\eta) = r_{{\rm{HP}},\min} = \min_\lambda r(\lambda,\eta=0).
 \label{Eq:HP:IC:HPline:RS}
\end{equation}
In other words, our algorithm with the parameters $\lambda$ and $\eta$ lying on the HP line has the same performance as the original HeatS+ProbS hybrid algorithm. When the parameter point $(\lambda,\eta)$ falls with the HP line such that $r(\lambda,\eta) < r_{{\rm{HP}},\min}$, our algorithm outperforms the HeatS+ProbS hybrid algorithm. On the contrary, the HeatS+ProbS hybrid algorithm performs better when the parameter point locates outside the HP line, that is, $r(\lambda,\eta) > r_{{\rm{HP}},\min}$.

For the MovieLens data, the ranking score reaches its minimum $r_{\min}=0.079$ when $\lambda=\lambda_{\rm{opt}}=0.26$ and $\eta=\eta_{\rm{opt}}=-0.71$. Compared with the minimal ranking score $r_{{\rm{HP}},\min}=0.840$ at the optimal $\lambda=\lambda_{\rm{HP,opt}}=0.16$ for the HeatS+ProbS hybrid algorithm, we gain an improvement of recommendation accuracy by $1-r_{\min}/r_{{\rm{HP}},\min}=6.0\%$. For the Netflix data, we have $r_{\min}=0.039$ when $\lambda_{\rm{opt}}=0.21$ and $\eta_{\rm{opt}}=-0.51$ and $r_{{\rm{HP}},\min}=0.045$ when $\lambda_{\rm{HP,opt}}=0.23$. We gain an improvement of recommendation accuracy by 12.8\%. It is found that $\lambda_{\rm{opt}}$ is close but not necessarily equal to $\lambda_{\rm{HP,opt}}$.

The second measure is the recommendation precision, which is defined as follows \cite{Zhou-Kuscsik-Liu-Medo-Wakeling-Zhang-2010-PNAS}
\begin{equation}
  P=\frac{1}{m}\frac{\sum_i^m d_{iL}}{L}
  \label{Eq:precision}
\end{equation}
where $d_{iL}$ is the number of user $u_i$'s deleted links contained in the top $L$ objects of his recommendation list. Plots (b) and (e) of fig.~\ref{Fig:HPIC:Accuracy} illustrate the contour lines of the precision functions $P(\lambda,\eta)$ with $L=50$ for the two data sets. Analogous to the HP line for ranking score, we can define the {\em{HP line}} for precision,
\begin{equation}
 P(\lambda,\eta) = P_{{\rm{HP}},\max} = \max_\lambda P(\lambda,\eta=0).
 \label{Eq:HP:IC:HPline:Prec}
\end{equation}
From fig.~\ref{Fig:HPIC:Accuracy}, we observe significant improvements achieved by our algorithm.

For the MovieLens data, we find that $P_{\max}=0.0904$ when $\lambda=\lambda_{\rm{opt}}=0.31$ and $\eta=\eta_{\rm{opt}}=-0.69$ in our algorithm, while $P_{{\rm{HP}},\max}=0.0865$ at the optimal $\lambda=\lambda_{\rm{HP,opt}}=0.30$ for the original HeatS+ProbS hybrid algorithm. We gain an improvement of recommendation accuracy by $P_{\max}/P_{{\rm{HP}},\max}-1=4.3\%$. For the Netflix data, we have $P_{\max}=0.0593$ when $\lambda_{\rm{opt}}=0.21$ and $\eta_{\rm{opt}}=-0.39$ and $P_{{\rm{HP}},\max}=0.0564$ when $\lambda_{\rm{HP,opt}}=0.20$. We gain an improvement of recommendation accuracy by 5.1\%.

The third measure is the recall, which is defined as follows \cite{Zhou-Kuscsik-Liu-Medo-Wakeling-Zhang-2010-PNAS}
\begin{equation}
 R=\frac{1}{m}\sum_i^m \frac{d_{iL}}{l_i}
 \label{Eq:recall}
\end{equation}
where $d_{iL}$ is the number of user $u_i$'s deleted links contained in the top $L$ objects, and $l_i$ is the number of user $u_i$'s deleted links. Plots (c) and (f) of fig.~\ref{Fig:HPIC:Accuracy} illustrate the contour lines of the precision functions $R(\lambda,\eta)$ with $L=50$ for the two data sets. Similarly, we can define the {\em{HP line}} for recall,
\begin{equation}
 R(\lambda,\eta) = R_{{\rm{HP}},\max} = \max_\lambda R(\lambda,\eta=0).
 \label{Eq:HP:IC:HPline:Recall}
\end{equation}
From fig.~\ref{Fig:HPIC:Accuracy}, we also observe significant improvements achieved by our algorithm.

For the MovieLens data, we have $R_{\max}=0.559$ when $\lambda=\lambda_{\rm{opt}}=0.31$ and $\eta=\eta_{\rm{opt}}=-0.51$ in our algorithm, while $R_{{\rm{HP}},\max}=0.548$ when $\lambda=\lambda_{\rm{HP,opt}}=0.29$ for the original HeatS+ProbS hybrid algorithm. We gain an improvement of recommendation accuracy by $R_{\max}/R_{{\rm{HP}},\max}-1=2.0\%$. For the Netflix data, we have $R_{\max}=0.439$ when $\lambda_{\rm{opt}}=0.21$ and $\eta_{\rm{opt}}=-0.29$ and $R_{{\rm{HP}},\max}=0.430$ when $\lambda_{\rm{HP,opt}}=0.21$. We gain an improvement of recommendation accuracy by 2.1\%.

\section{Diversity of recommendation}

We adopt two measures to characterize the diversity of recommendations, the intra-user diversity $D_{\rm{intra}}$ and the inter-user diversity $D_{\rm{inter}}$.

The intra-user diversity characterizes the average dissimilarity among the top $L$ objects in a single user's list, denoted $\mathcal{L}_i$. The similarity between two objects $o_\alpha$ and $o_\beta$ can be measured by the S{\o}rensen index \cite{Sorensen-1948-BK}
\begin{equation}
  s_{\alpha\beta}=\frac{1}{\sqrt{k_\alpha k_\beta}}\sum_{i=1}^ma_{i\alpha}a_{i\beta},
  \label{Eq:SorensenInd}
\end{equation}
and the intra-user diversity of user $u_i$'s recommendation list of length $L$ can be defined as \cite{Ziegler-McNee-Konstan-Lausen-2005-WWW,Zhou-Su-Liu-Jiang-Wang-Zhang-2009-NJP}
\begin{equation}
 D^i_{\rm{intra}} = \frac{1}{L(L-1)}\sum_{\alpha \neq \beta} (1-s_{\alpha \beta}),
 \label{Eq:IntraDiversity:i}
\end{equation}
where $o_\alpha\in\mathcal{L}_i$ and $o_\beta\in\mathcal{L}_i$, and the average intra-user diversity is
\begin{equation}
 D_{\rm{intra}} = \frac{1}{m}\sum_{i=1}^m D^i_{\rm{intra}}.
 \label{Eq:IntraDiversity}
\end{equation}
Note that $\mathcal{L}_i$ and $\mathcal{L}_j$ of any two users are usually different and thus their $D^i_{\rm{intra}}$ and $D^j_{\rm{intra}}$ values differ from one user to another. A greater or lesser value of the intra-user diversity means higher or lower novelty of a single user's recommendation list.

The inter-user diversity indicates the uniqueness of different users' recommendation lists, which can be calculated as follows \cite{Zhou-Su-Liu-Jiang-Wang-Zhang-2009-NJP,Zhou-Kuscsik-Liu-Medo-Wakeling-Zhang-2010-PNAS}
\begin{equation}
 D_{\rm{inter}}=\frac{2}{m(m-1)} \sum_{j=1}^m\sum_{i=j+1}^m\left(1-\frac{|\mathcal{L}_i\cap\mathcal{L}_j|}{L}\right)
 \label{Eq:InterDiversity}
\end{equation}
where $|\mathcal{L}_i\cap\mathcal{L}_j|$ is the number of the common objects of the top $L$ of the two lists $\mathcal{L}_i$ and $\mathcal{L}_j$. The inter-user diversity reflects the uniqueness of different users' recommendation lists and is a measure of personalization of the recommendation algorithm. A greater or lesser value of the inter-user diversity means higher or lower personalization of users' recommendation lists.

\begin{figure}[htb]
 \centering
 \includegraphics[width=4cm]{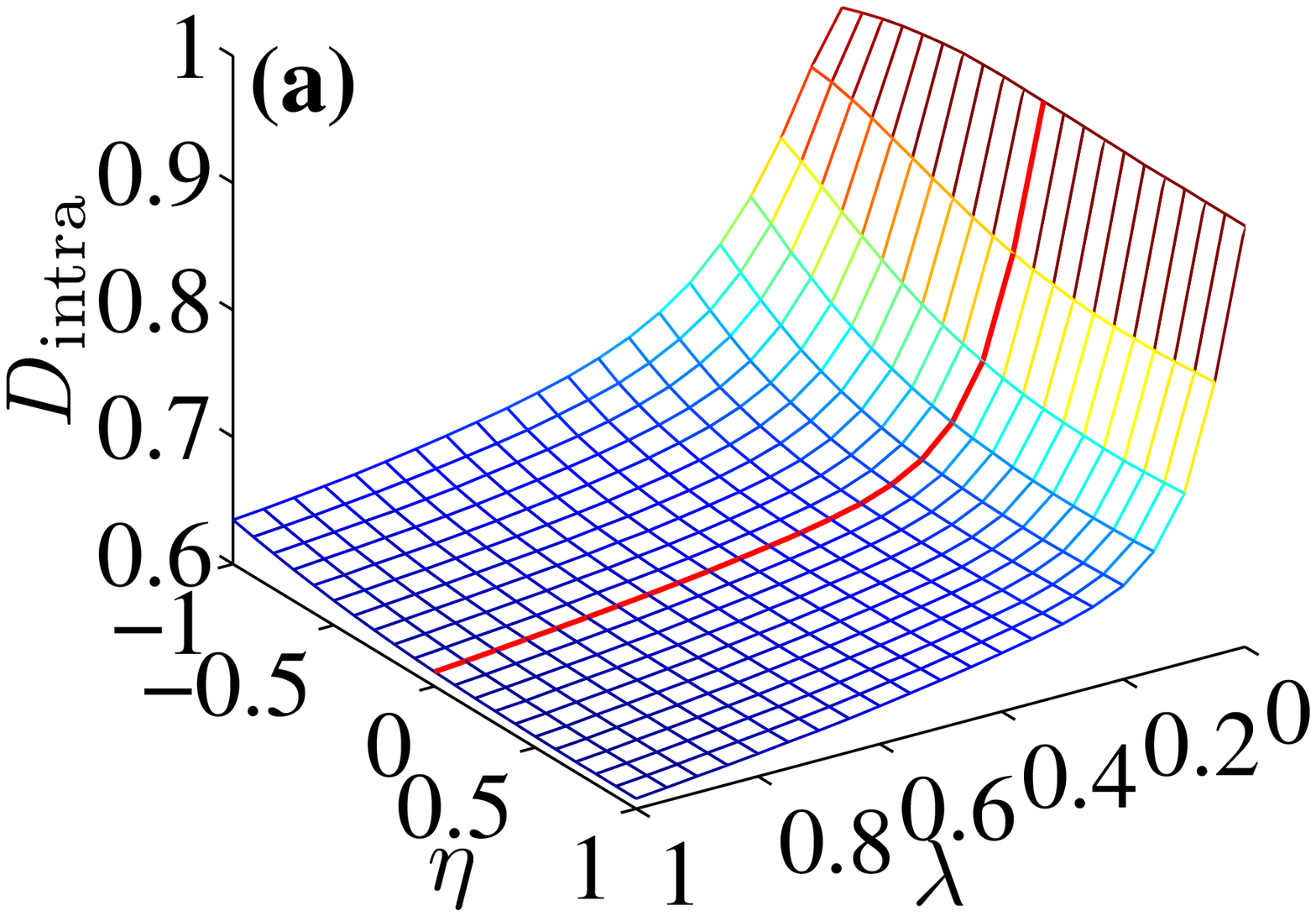}
 \includegraphics[width=4cm]{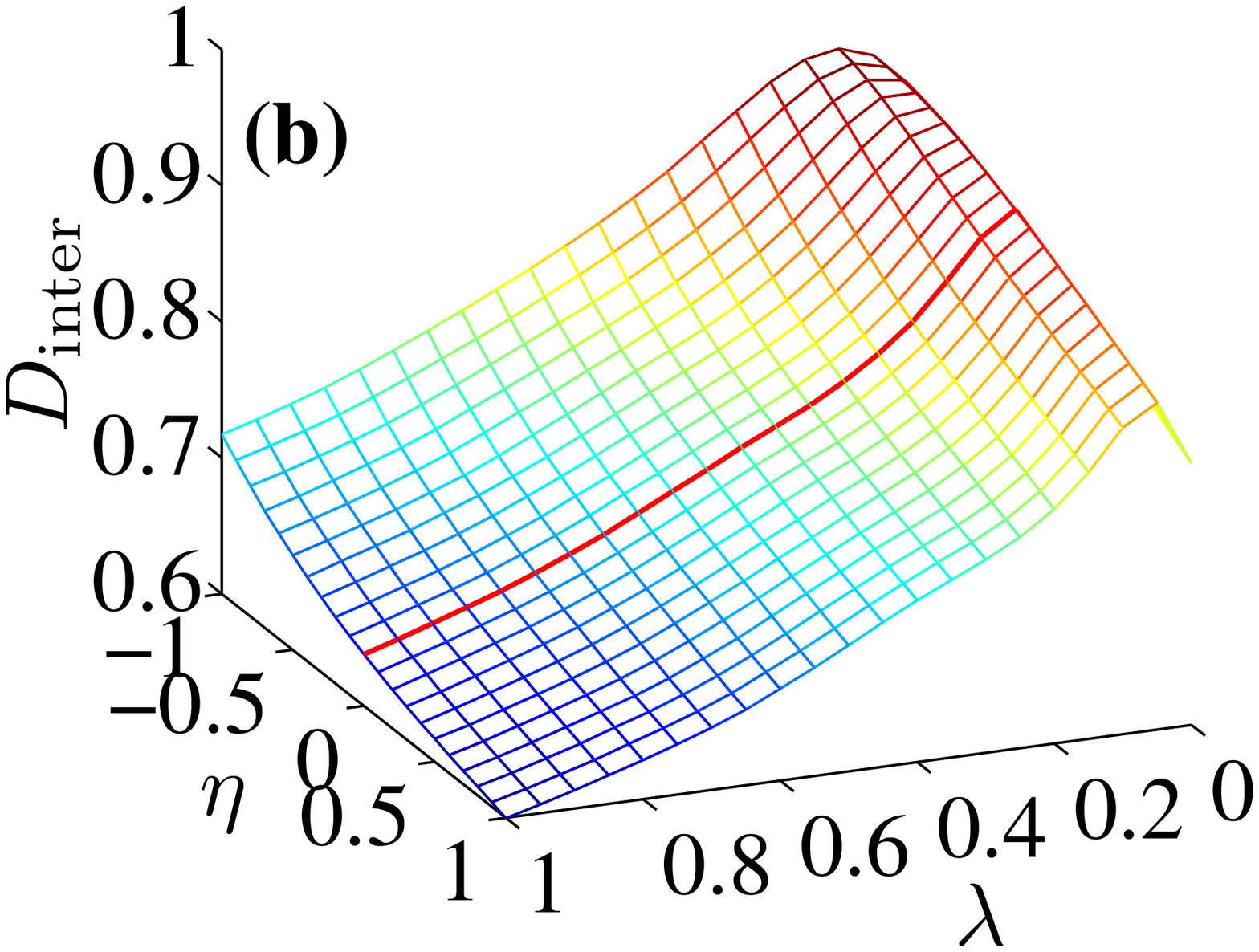}
 \includegraphics[width=4cm]{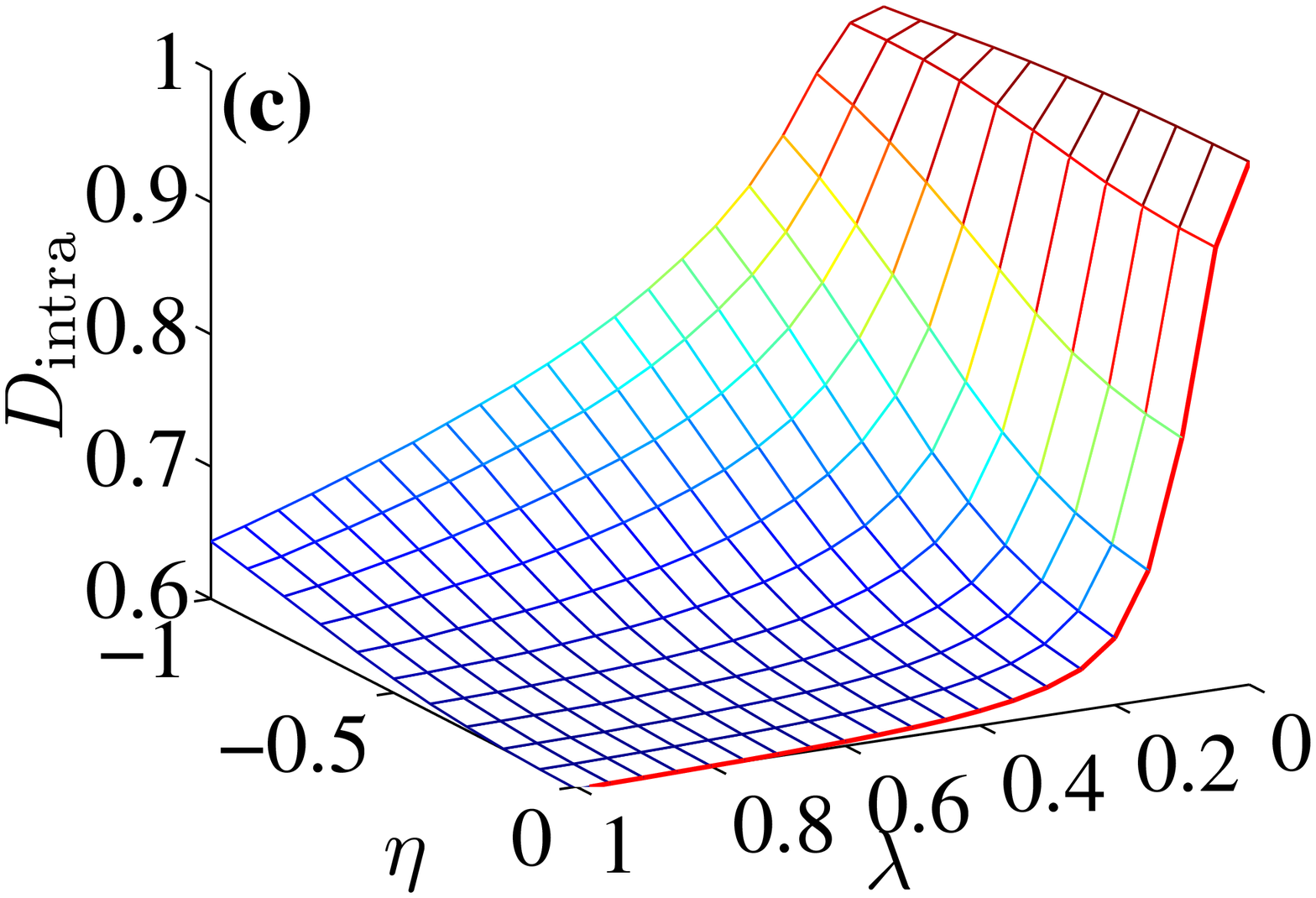}
 \includegraphics[width=4cm]{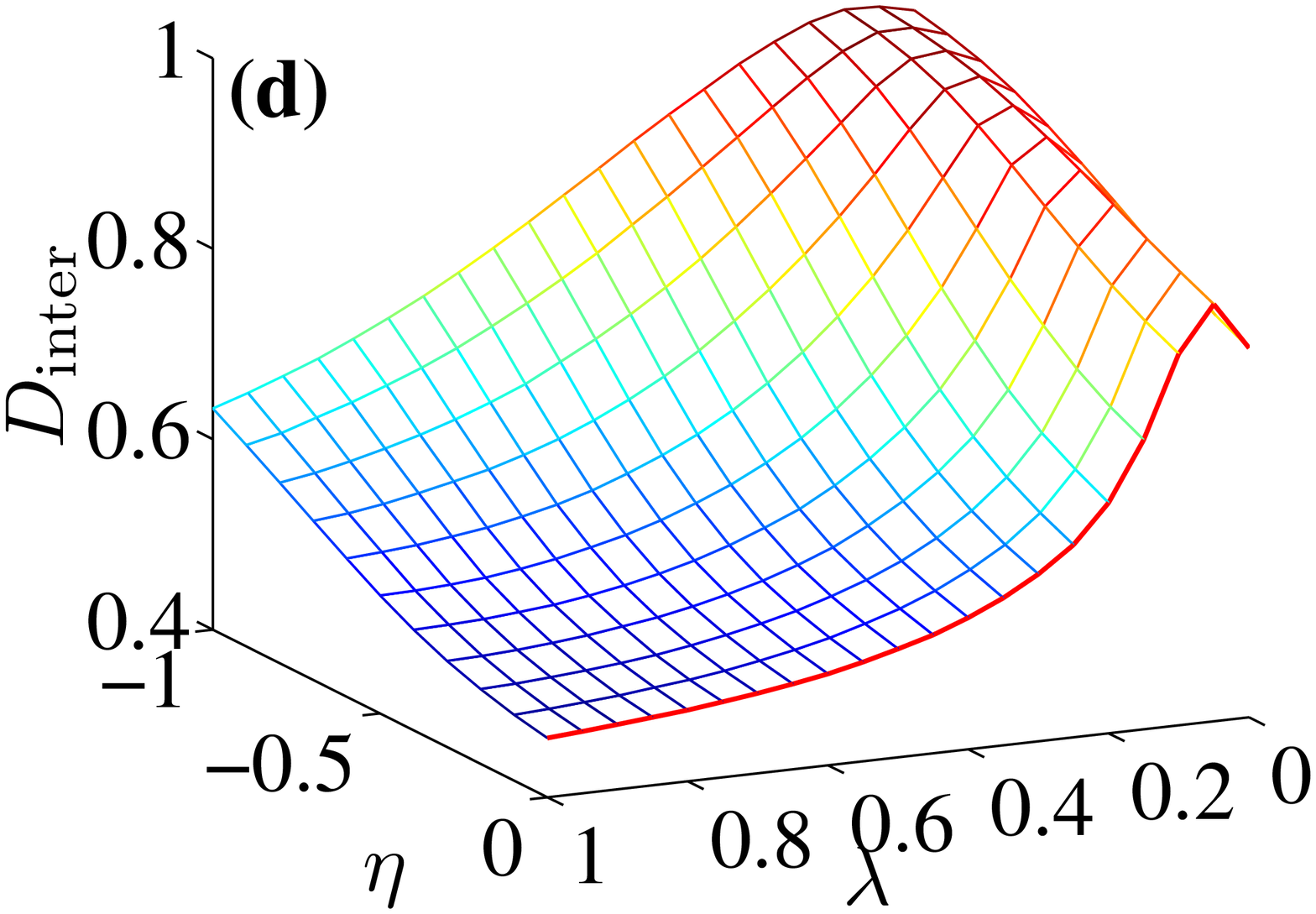}
 \caption{\label{Fig:HPIC:Diversity} (Color online) Intra-user diversity (a, c) and inter-user diversity (b, d) as a function of $\lambda$ and $\eta$ for MovieLens (a, b) and Netflix (c, d). The length of the recommendation list is $L=50$. The results for the original HeatS+ProbS hybrid algorithm are given with $\eta=0$.}
\end{figure}

Figure \ref{Fig:HPIC:Diversity} shows the dependence of the intra-user diversity and the inter-user diversity as a function of $\lambda$ and $\eta$ for the MovieLens and Netflix datasets. The length of the recommendation list is $L=50$. The four plots share several similar features. For a fixed value of $\eta$, the diversity decreases with increasing $\lambda$. This is expected since the HeatS part of the hybrid algorithm dominates when $\lambda$ is small. For a fixed value of $\lambda$, the diversity decreases with increasing $\eta$, which can be understood that, for larger values of $\eta$, more resource is put on the objects with larger degrees and the algorithm favors the ProbS part. Combined with the results in fig.~\ref{Fig:HPIC:Accuracy}, our algorithm with negative $\eta$ values can improve both the accuracy and diversity of the recommendation. This finding is consistent with the results in ref.~\cite{Zhou-Jiang-Su-Zhang-2008-EPL}, in which $\lambda=1$.

\section{Performance comparison of recommendation algorithms}

We compare the performance of four recommendation algorithms, HeatS, ProbS, HeatS+ProbS, and HeatS+ProbS with heterogeneous initial configuration (HPIC). The comparison is based on three accuracy measures (ranking score $r$, precision $P$, and recall $R$) and two diversity measures (intra-user diversity $D_{\rm{intra}}$ and inter-user diversity $D_{\rm{inter}}$) using the MovieLens and Netflix data sets, respectively. The parameter $\lambda$ of the HeatS+ProbS hybrid algorithm is tuned to minimize the ranking score, and so are the two parameters of the HPIC algorithm. For the MovieLens data, we have $\lambda_{\rm{HP,opt}} = 0.16$ for the HeatS+ProbS algorithm and $\lambda_{\rm{opt}}=0.26$ and $\eta_{\rm{opt}}=-0.71$ for the HPIC algorithm. For the Netflix data, we have $\lambda_{\rm{HP,opt}} = 0.23$ for the HeatS+ProbS algorithm and $\lambda_{\rm{opt}}=0.21$ and $\eta_{\rm{opt}}=-0.51$ for the HPIC algorithm.

Table \ref{TB:PerformanceComp} shows the results. For both data sets, the ranking score $r$ decreases, and  both the precision $P$ and the recall $R$ increase from left to right, except that the recall of HPIC is smaller than that of HeatS+ProbS. It means that the recommendation accuracy improves from HeatS to ProbS to HeatS+ProbS to HPIC. Concerning the recommendation diversity, the HeatS algorithm gives the largest diversity values and the ProbS algorithm results in the smallest diversity values. The improvement of recommendation diversity after introducing heterogeneous initial configuration in the HeatS+ProbS hybrid algorithm is marginal for the MovieLens data. However, we can observe a significant increase in the two diversity measures for the Netflix data. Therefore, we can conclude that introducing heterogeneous initial configuration in the HeatS+ProbS hybrid algorithm can remarkably improve the recommendation accuracy and increase more or less the recommendation diversity.

\begin{table}[htbp]
 \centering
 \caption{\label{TB:PerformanceComp} Performance comparison of different recommendation algorithms according to each of the five metrics: ranking score $r$, precision $P$, recall $R$, intra-user diversity $D_{\rm{intra}}$, and inter-user diversity $D_{\rm{inter}}$. For the MovieLens data, $\lambda_{\rm{HP,opt}} = 0.16$ for the HeatS+ProbS algorithm and $\lambda_{\rm{opt}}=0.26$ and $\eta_{\rm{opt}}=-0.71$ for our algorithm (HPIC). For the Netflix data, $\lambda_{\rm{HP,opt}} = 0.23$ for the HeatS+ProbS algorithm and $\lambda_{\rm{opt}}=0.21$ and $\eta_{\rm{opt}}=-0.51$ for our algorithm.}
 \medskip
 \begin{tabular}{ccccc}
  \hline\hline
   & HeatS & ProbS & HeatS+ProbS & HPIC \\ \hline
  \multicolumn{5}{l}{MovieLens}\\\hline
  $r$ & 0.149 & 0.106 & 0.084 & 0.079\\
  $P$ & 0.023 & 0.074 & 0.084 & 0.089\\
  $R$ & 0.130 & 0.476 & 0.501 &  0.544    \\
  $D_{\rm{intra}}$ &  0.932 & 0.638 & 0.699 & 0.694   \\
  $D_{\rm{inter}}$ &  0.862 & 0.618 & 0.853 & 0.867 \\
  \hline
  \multicolumn{5}{l}{Netflix}\\\hline
  $r$ & 0.107 & 0.050 & 0.045 & 0.039\\
  $P$ & 0.014 & 0.050 & 0.056 & 0.059\\
  $R$ & 0.022 & 0.385 & 0.429 & 0.426     \\
  $D_{\rm{intra}}$ & 0.995 & 0.598 & 0.641  &  0.721 \\
  $D_{\rm{inter}}$ & 0.788 & 0.462 & 0.624  &  0.780 \\
  \hline\hline
   \end{tabular}
\end{table}

\section{Dependence of algorithm accuracy on the object degree}

\begin{figure*}[htb]
\centering
\includegraphics[width=4.3cm]{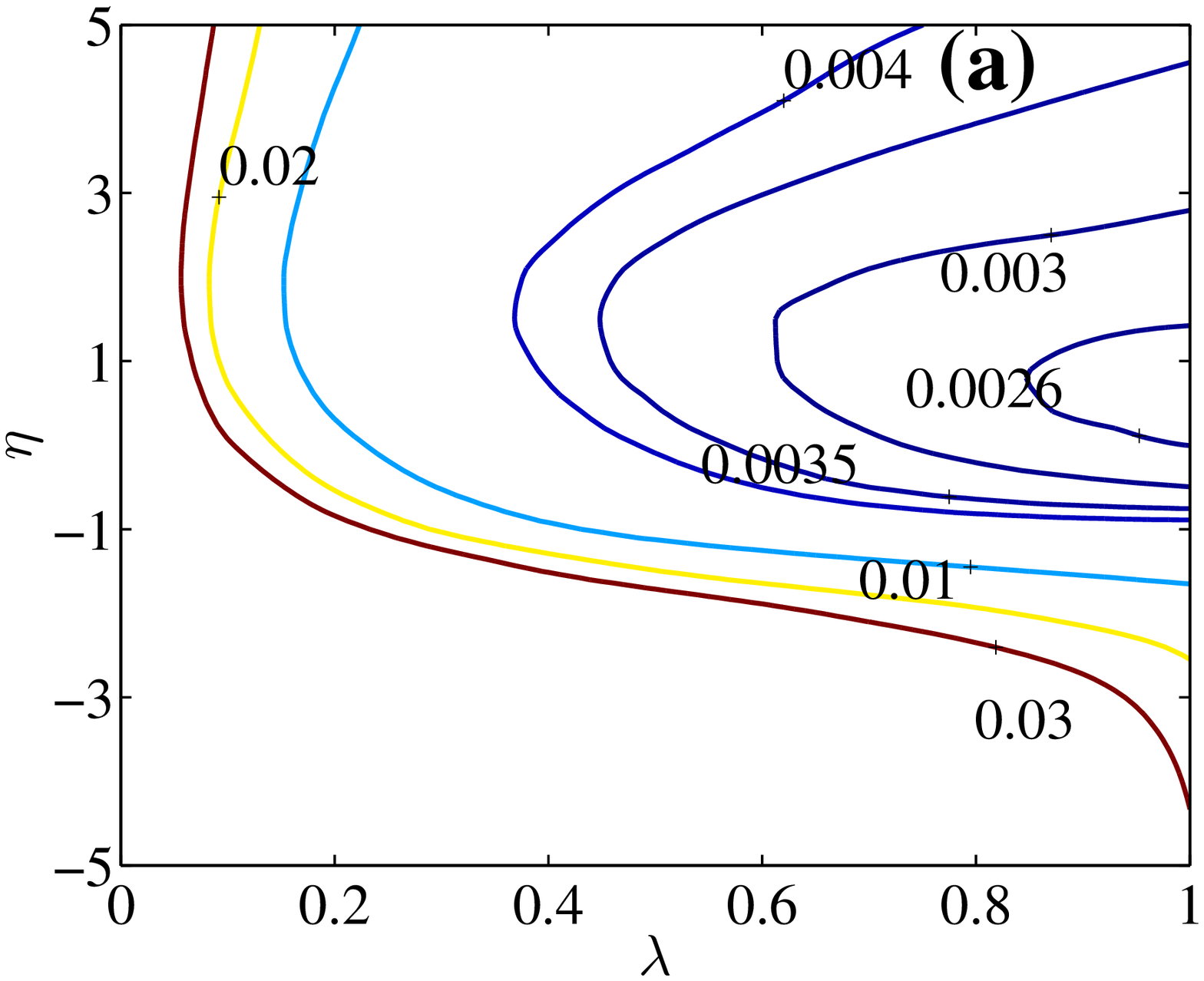}
\includegraphics[width=4.3cm]{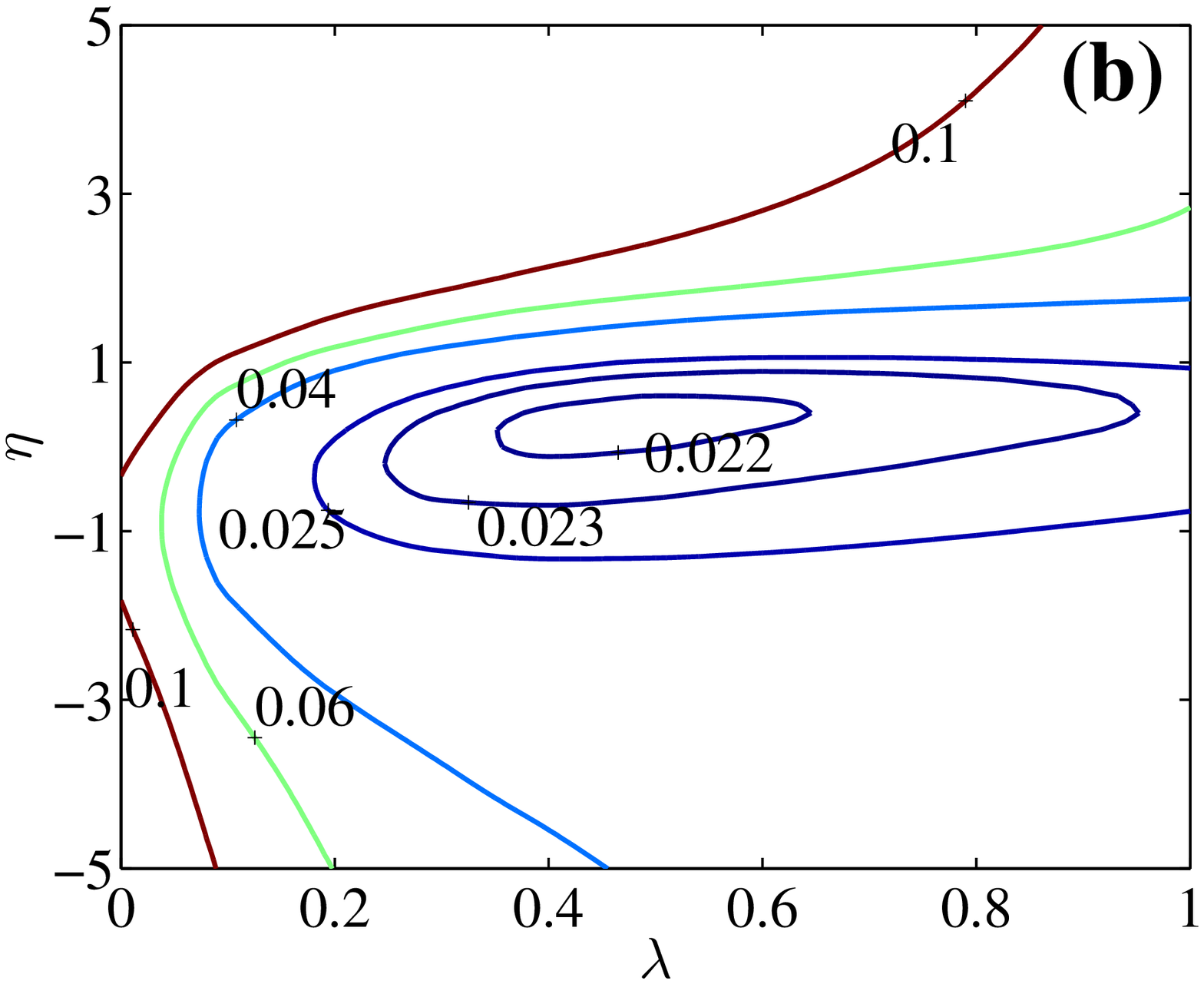}
\includegraphics[width=4.3cm]{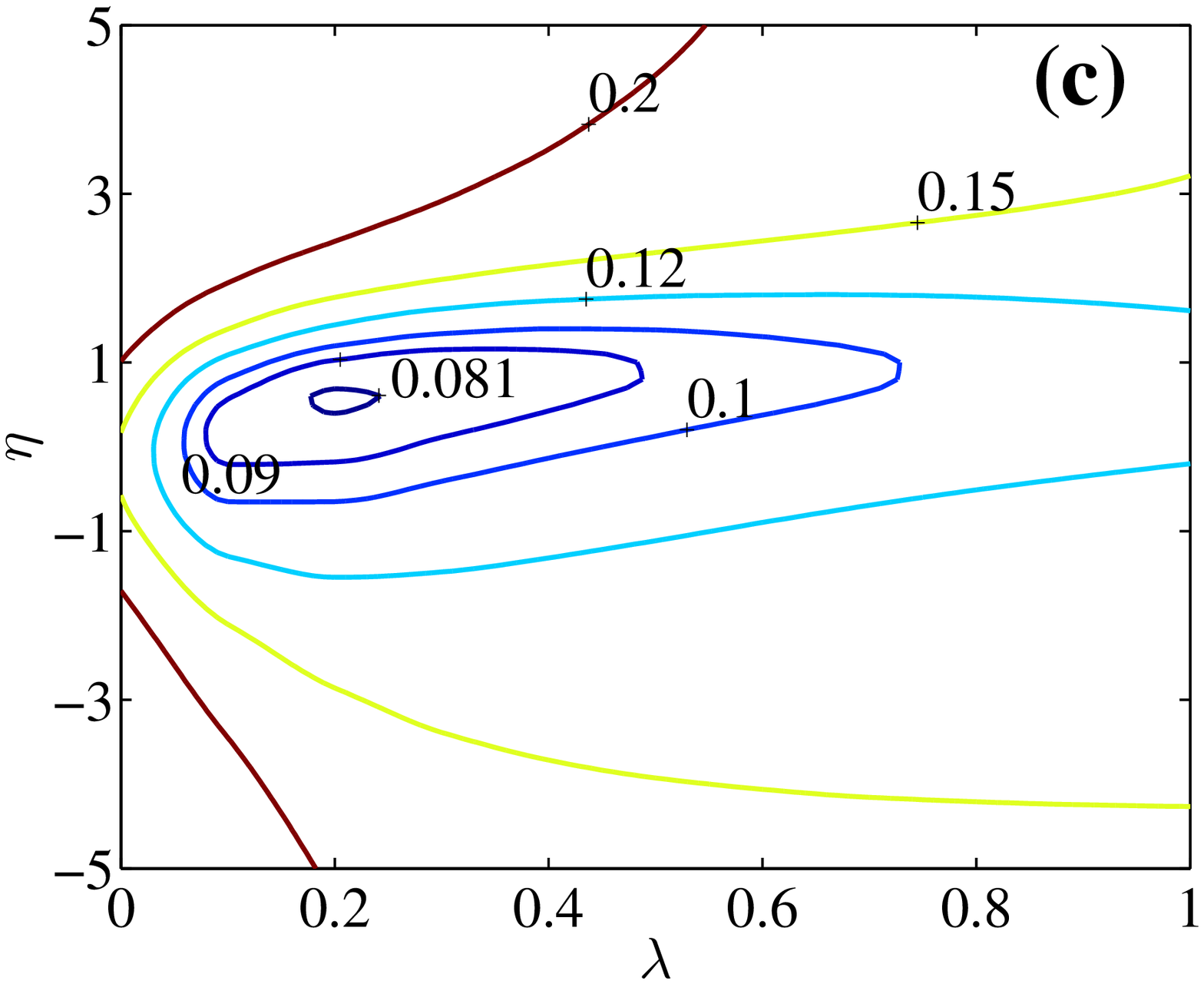}
\includegraphics[width=4.3cm]{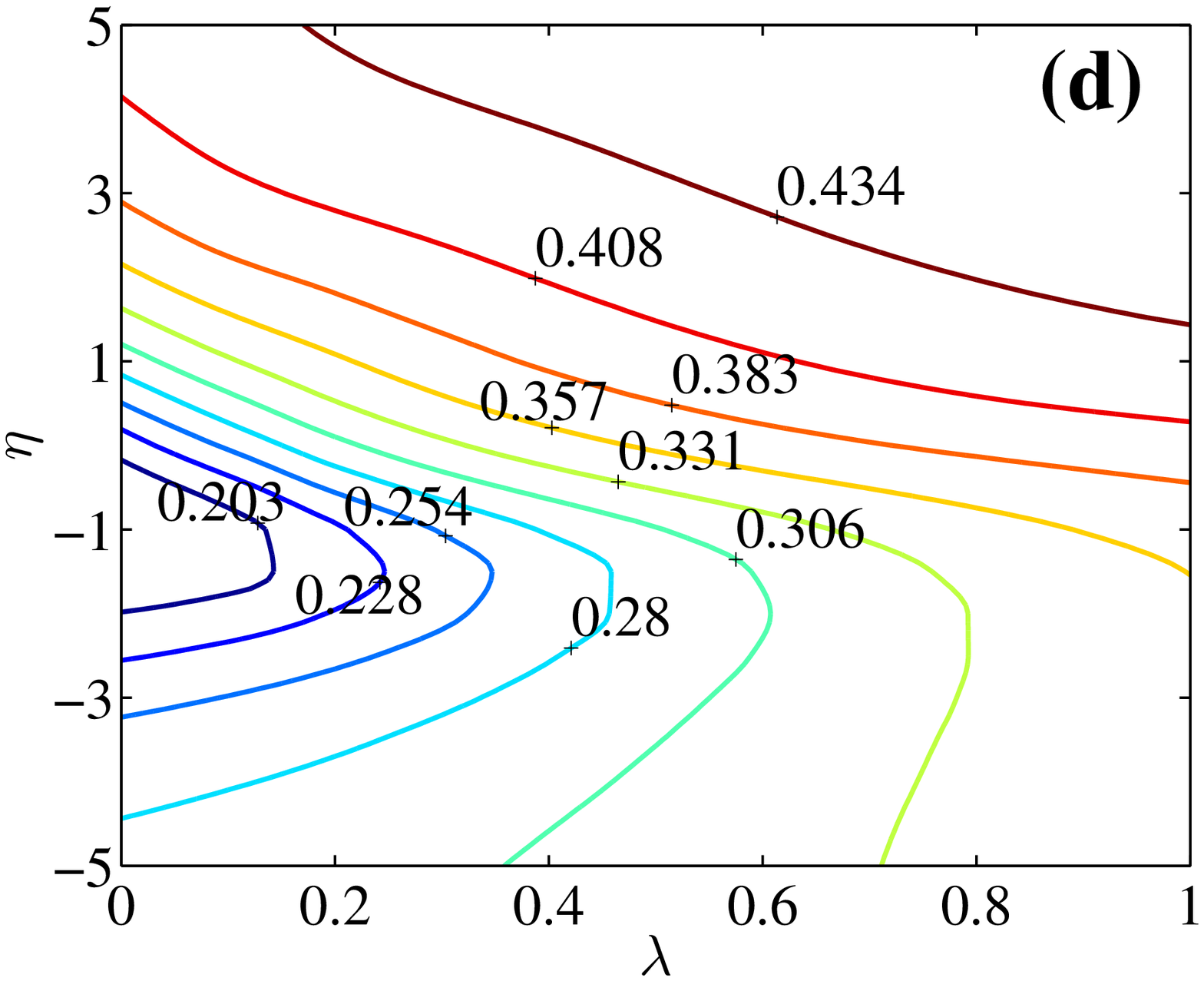}
\includegraphics[width=4.3cm]{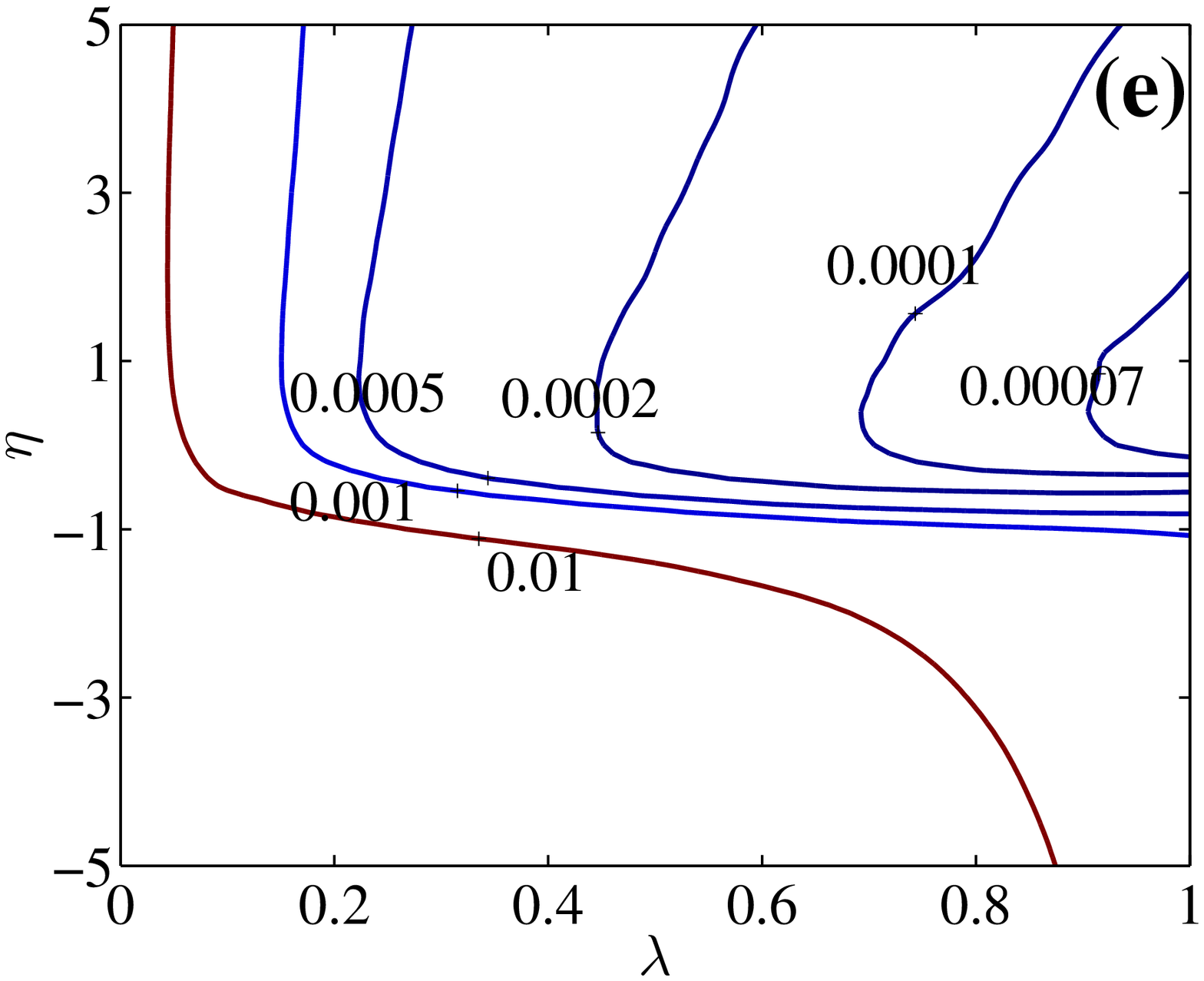}
\includegraphics[width=4.3cm]{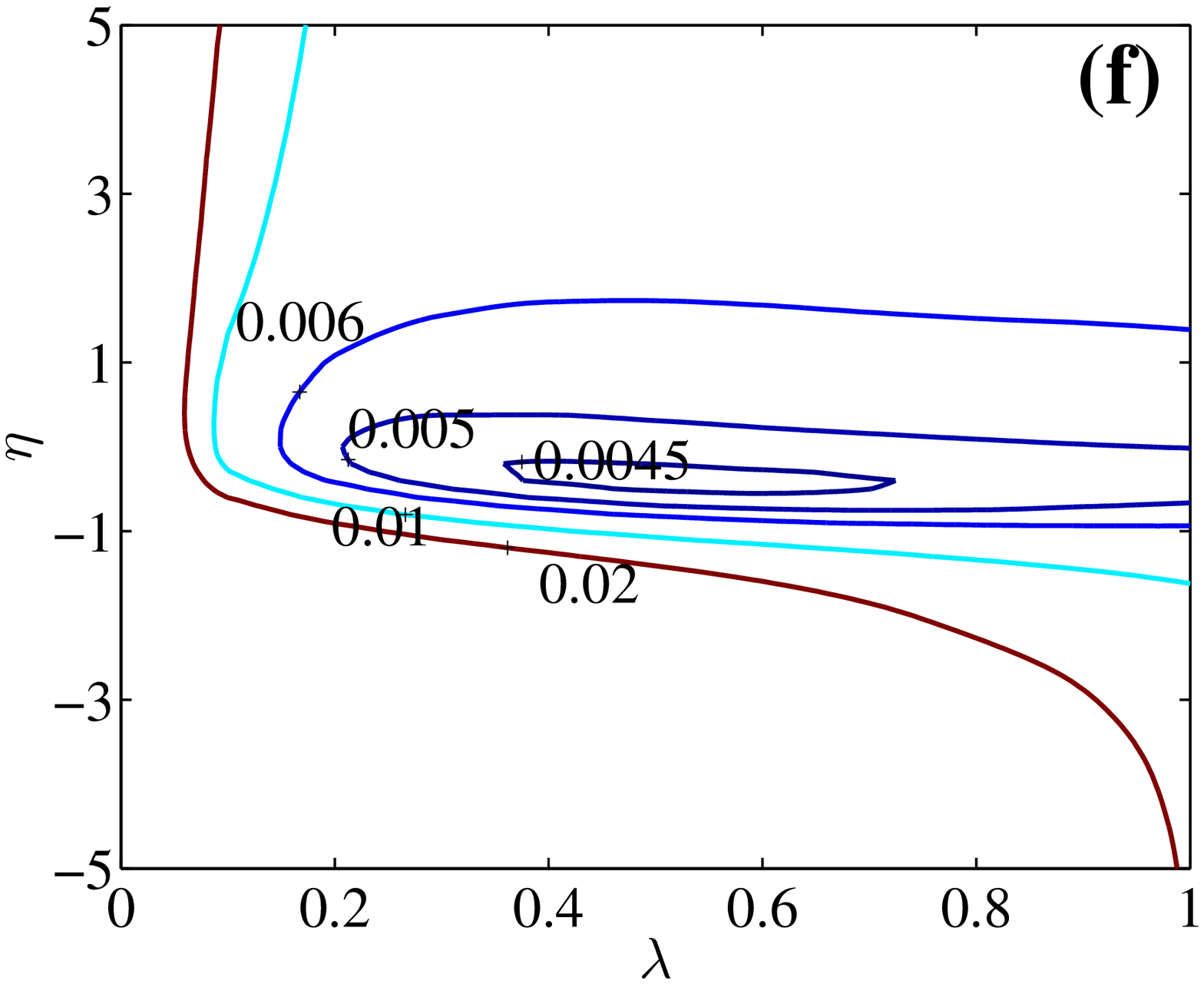}
\includegraphics[width=4.3cm]{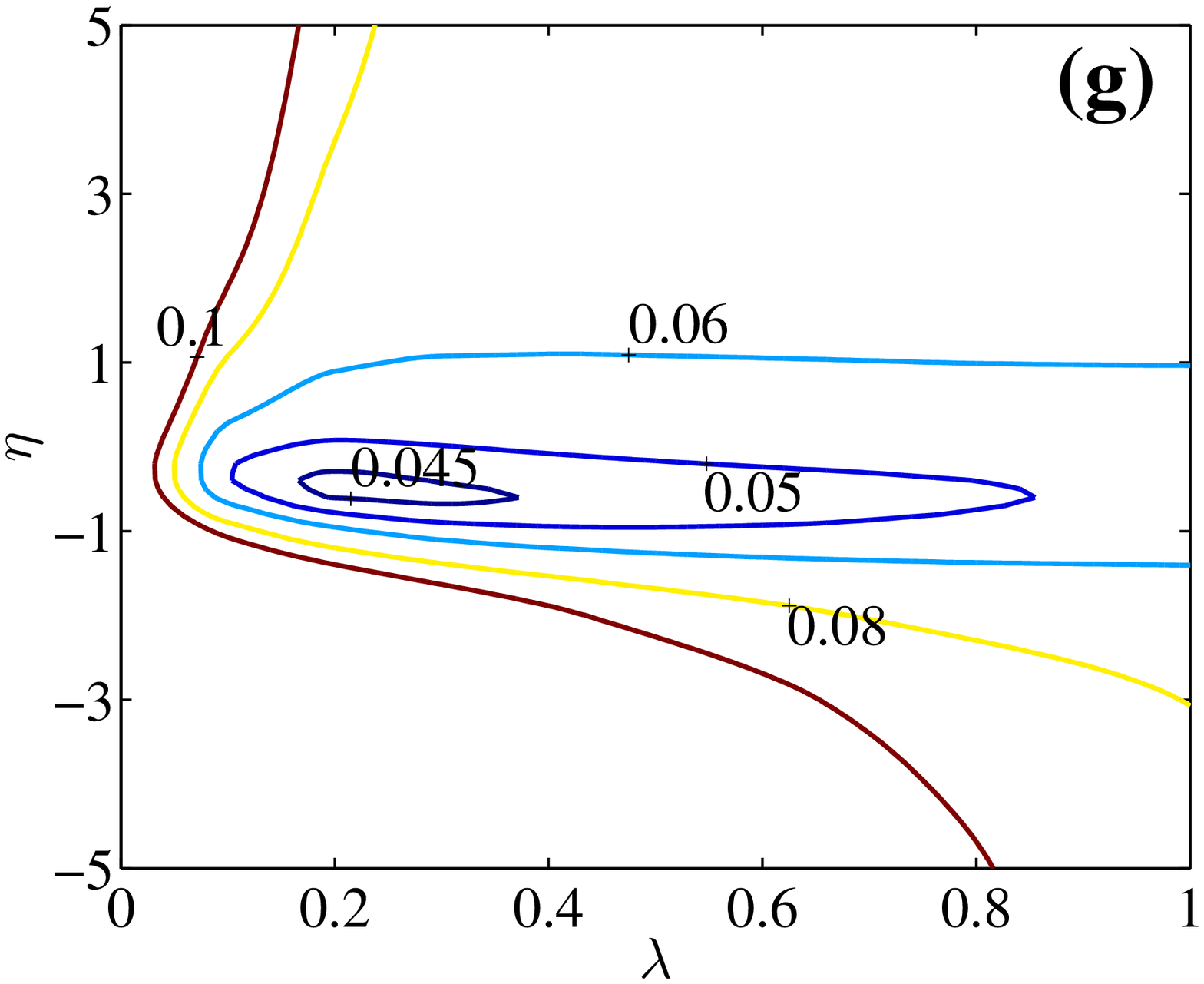}
\includegraphics[width=4.3cm]{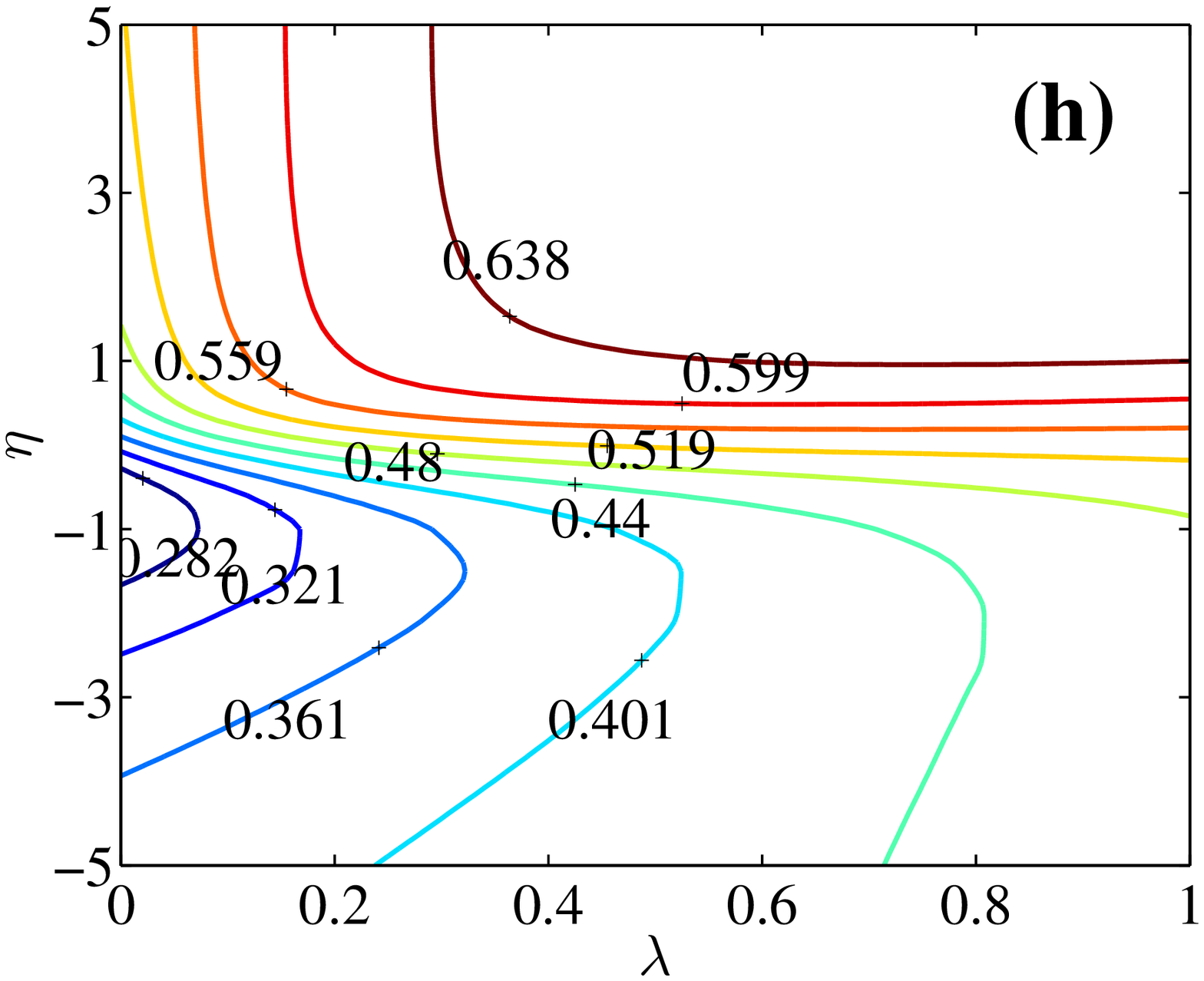}
\caption{\label{Fig:HPIC:PopObj} (Color online) Contour plots of the ranking score $r(\lambda,\eta)$ for four circumstances by averaging the ranking scores of 1000 entries with different object degrees for MovieLens (a-d) and Netflix (e-h). The object degree decreases from left to right.}
\end{figure*}

The above investigation focuses on the macroscopic performance of the recommendation algorithms. It will be helpful to understand the recommendation algorithm at the microscopic level by studying the dependence of algorithm accuracy on the object degree \cite{Zhou-Jiang-Su-Zhang-2008-EPL,Zhou-Kuscsik-Liu-Medo-Wakeling-Zhang-2010-PNAS}. In doing so, the entries in each 10\% probe set are sorted according to a descending order of object degrees. Four new probe sets, each containing 1000 links, are extracted from the 10\% probe set: the most popular objects with the highest degrees, popular objects with high degrees starting from one fourth of the sorted link sequence, unpopular objects with low degrees starting from the middle of the sorted sequence, and the least popular objects with the lowest degrees, respectively. The average ranking scores corresponding to these four probe sets are calculated for the HPIC algorithm with different values of the two parameters. The results for popular and unpopular objects are illustrated in fig.~\ref{Fig:HPIC:PopObj} with contours.

For popular objects, according to fig.~\ref{Fig:HPIC:PopObj}(a) and (e), the ranking score is negatively correlated with $\lambda$ for fixed $\eta$ and the minimal ranking score is reached at $\lambda\approx1$ and $\eta\approx0.5$. The recommendation performance remains good when $\lambda$ is large and $\eta$ is positive, which is particularly evident for the Netflix data, see fig.~\ref{Fig:HPIC:PopObj}(e). This observation is consistent with the fact that popular objects are more likely to be recommended when the ProbS algorithm dominates and/or popular objects are configured with more initial resources ($\eta>0$). For unpopular objects, according to fig.~\ref{Fig:HPIC:PopObj}(d) and (h), the ranking score is positively correlated with $\lambda$ for fixed $\eta$ and the minimal ranking score is reached at $\lambda\approx0$ and $\eta\approx-1$. This finding is consistent with the fact that unpopular objects become more likely to be recommended when the HeatS algorithm dominates and/or unpopular objects are configured with more initial resources ($\eta<0$).

Comparing the plots from fig.~\ref{Fig:HPIC:PopObj}(a) to fig.~\ref{Fig:HPIC:PopObj}(d) for MovieLens or from fig.~\ref{Fig:HPIC:PopObj}(e) to fig.~\ref{Fig:HPIC:PopObj}(h) for Netflix, the overall ranking score decreases with the object degree, which is expected since popular objects are more frequently collected by users that makes them popular. For each data set, the optimal point $(\lambda_{\rm{opt}},\eta_{\rm{opt}})$ corresponding to the minimum ranking score in the investigated region $(\lambda,\eta)=[0,1]\times[-5,5]$ moves from northeast to southwest when the object degree increases. The optimal parameter values $\lambda_{\rm{opt}}$ and $\eta_{\rm{opt}}$ are listed in table \ref{TB:ko}.

\begin{table}[htb]
 \centering
 \caption{\label{TB:ko} Optimal parameter values $\lambda_{\rm{opt}}$ and $\eta_{\rm{opt}}$ in the investigated region $(\lambda,\eta)=[0,1]\times[-5,5]$ for four probe sets with different object degrees.}
 \medskip
 \begin{tabular}{ccccccccccccccc}
  \hline\hline
   \multirow{3}*[2mm]{Object degree} & \multicolumn{2}{c}{MovieLens} && \multicolumn{2}{c}{Netflix}\\
   \cline{2-3}\cline{5-6}
   & $\lambda_{\rm{opt}}$ & $\eta_{\rm{opt}}$ && $\lambda_{\rm{opt}}$ & $\eta_{\rm{opt}}$ \\ \hline
  Highest  & 1.0 & $~0.4$ && 1.0 & $~0.5$ \\
  High     & 0.5 & -0.4 && 0.5 & -0.3 \\
  Low      & 0.2 & -0.6 && 0.2 & -0.4    \\
  Lowest   & 0.0 & -1.0 && 0.0 & -1.0   \\
  \hline\hline
   \end{tabular}
\end{table}

\section{\label{S1:Concl}Conclusion}

In this work, we have proposed to use heterogeneous initial resource configuration in the HeatS+ProbS hybrid recommendation algorithm. An additional parameter $\eta$ is introduced in this algorithm. We investigated the recommendation performance using three accuracy measures and two diversity measures testes on two benchmark data sets, MovieLens and Netflix. Numerical experiments indicate that assigning less initial resource on popular objects and more initial resource on unpopular objects provides systematic improvements in all these measures. More interestingly from the practical point of view, our algorithm is robust since the parameter region enclosed by the so-called HP line is broad.

In order to understand the behavior of the proposed recommender system on the microscopic level, we investigated the recommendation accuracy of objects with different degrees. We found that the recommendation accuracy is sensitive to both parameters. Popular objects with high degrees have higher recommendation accuracy when the ProbS part dominates ($\lambda=1$) and popular objects are assigned with more initial resource ($\eta>0$), while unpopular objects are more accurately recommended when the HeatS part dominates ($\lambda=0$) and popular objects are assigned with less initial resource ($\eta<0$).

In summary, introducing heterogeneity in the initial configuration of resource on objects can improve the recommendation performance of the HeatS+ProbS hybrid algorithm, which is the best network-based recommendation algorithm to date in which both accuracy and diversity are taken into consideration. The complexity of recommender systems uncovered in this work highlights the possibility of further improvements in algorithm design.

\acknowledgments

This work was partly supported by the Program for New Century Excellent Talents in University under Grant No. NCET-07-0288.

\bibliography{E:/Papers/Auxiliary/Bibliography}

\begin{thebibliography}{10}
\expandafter\ifx\csname url\endcsname\relax\def\url#1{\texttt{#1}}\fi

\bibitem{Albert-Barabasi-2002-RMP}
\Name{Albert R. \and Barab{\'a}si A.-L.} \REVIEW{Rev. Mod. Phys.
  }{74}{2002}{47}.

\bibitem{Newman-2003-SIAMR}
\Name{Newman M. E.~J.} \REVIEW{SIAM Rev. }{45}{2003}{167}.

\bibitem{Boccaletti-Latora-Moreno-Chavez-Hwang-2006-PR}
\Name{Boccaletti S., Latora V., Moreno Y., Chavez M. \and Hwang D.-U.}
  \REVIEW{Phys. Rep. }{424}{2006}{175}.

\bibitem{Sarukkai-2000-CN}
\Name{Sarukkai R.~R.} \REVIEW{Comput. Networks }{33}{2000}{377}.

\bibitem{LibenNowell-Kleinberg-2007-JASIST}
\Name{Liben-Nowell D. \and Kleinberg J.} \REVIEW{J. Am. Soc. Inf. Sci. Technol.
  }{58}{2007}{1019}.

\bibitem{Clauset-Moore-Newman-2008-Nature}
\Name{Clauset A., Moore C. \and Newman M. E.~J.} \REVIEW{Nature
  }{453}{2008}{98}.

\bibitem{Zhu-2001-LNAI}
\Name{Zhu J.-H.} \REVIEW{Lect. Note. Artif. Int. }{2109}{2001}{298}.

\bibitem{Zhu-Hong-Hughes-2002-LNCS}
\Name{Zhu J.-H., Hong J. \and Hughes J.-G.} \REVIEW{Lect. Notes Comput. Sci.
  }{2311}{2002}{60}.

\bibitem{Marchette-Priebe-2008-CSDA}
\Name{Marchette D.~J. \and Priebe C.~E.} \REVIEW{Comput. Statist. Data Anal.
  }{52}{2008}{1373}.

\bibitem{Zhou-Lu-Zhang-2009-EPJB}
\Name{Zhou T., Lu L.-Y. \and Zhang Y.-C.} \REVIEW{Eur. Phys. J. B
  }{71}{2009}{623}.

\bibitem{Lu-Jin-Zhou-2009-PRE}
\Name{Lu L.-Y., Jin C.-H. \and Zhou T.} \REVIEW{Phys. Rev. E
  }{80}{2009}{046122}.

\bibitem{Lu-Zhou-2010-EPL}
\Name{Lu L.-Y. \and Zhou T.} \REVIEW{EPL (Europhys. Lett.) }{89}{2010}{18001}.

\bibitem{Liu-Lu-2010-EPL}
\Name{Liu W.-P. \and Lu L.-Y.} \REVIEW{EPL (Europhys. Lett.)
  }{89}{2010}{58007}.

\bibitem{Resnick-Varian-1997-CACM}
\Name{Resnick P. \and Varian H.~R.} \REVIEW{Commun. ACM }{40}{1997}{56}.

\bibitem{Yildirim-Goh-Cusick-Barabasi-Vidal-2007-NBt}
\Name{Yildirim M.~A., Goh K.~I., Cusick M.~E., Barab{\'a}si A.-L. \and Vidal
  M.} \REVIEW{Nat. Biotechnol. }{25}{2007}{1119}.

\bibitem{Goldberg-Nichols-Oki-Terry-1992-CACM}
\Name{Goldberg D., Nichols D., Oki B.~M. \and Terry D.} \REVIEW{Commun. ACM
  }{35}{1992}{61}.

\bibitem{Herlocker-Konstan-Terveen-Riedl-2004-ACMtis}
\Name{Herlocker J.~L., Konstan J.~A., Terveen K. \and Riedl J.~T.} \REVIEW{ACM
  Trans. Info. Sys. }{22}{2004}{5}.

\bibitem{Burke-2002-UMUAI}
\Name{Burke R.} \REVIEW{User Model. User-Adapt. Inter. }{12}{2001}{331}.

\bibitem{Zhou-Ren-Medo-Zhang-2007-PRE}
\Name{Zhou T., Ren J., Medo M. \and Zhang Y.-C.} \REVIEW{Phys. Rev. E
  }{76}{2007}{046115}.

\bibitem{Jia-Liu-Sun-Wang-2008-PA}
\Name{Jia C.-X., Liu R.-R., Sun D. \and Wang B.-H.} \REVIEW{Physica A
  }{387}{2008}{5887}.

\bibitem{Zhou-Su-Liu-Jiang-Wang-Zhang-2009-NJP}
\Name{Zhou T., Su R.-Q., Liu R.-R., Jiang L.-L., Wang B.-H. \and Zhang Y.-C.}
  \REVIEW{New J. Phys. }{11}{2009}{123008}.

\bibitem{Liu-Deng-2009-PA}
\Name{Liu J. \and Deng G.-S.} \REVIEW{Physica A }{388}{2009}{3643}.

\bibitem{Zhou-Kuscsik-Liu-Medo-Wakeling-Zhang-2010-PNAS}
\Name{Zhou T., Kuscsik Z., Liu J.-G., Medo M., Wakeling J.~R. \and Zhang Y.-C.}
  \REVIEW{Proc. Natl. Acad. Sci. U.S.A. }{107}{2010}{4511}.

\bibitem{Zhou-Jiang-Su-Zhang-2008-EPL}
\Name{Zhou T., Jiang L.-L., Su R.-Q. \and Zhang Y.-C.} \REVIEW{EPL (Europhys.
  Lett.) }{81}{2008}{58004}.

\bibitem{Konstan-Miller-Maltz-Herlocker-Gordon-Riedl-1997-CACM}
\Name{Konstan J., Miller B., Maltz D., Herlocker J., Gordon L. \and Riedl J.}
  \REVIEW{Commun. ACM }{40}{1997}{77}.

\bibitem{Bennett-Lanning-2007-Netflix}
\Name{Bennett J. \and Lanning S.} \Book{{The Netflix Prize}} in
  \Book{Proceedings of the KDD Cup Workshop 2007} (ACM, New York) 2007 pp.
  3--6.

\bibitem{Sorensen-1948-BK}
\Name{S{\o}rensen T.} \REVIEW{Biol. Skr. }{5}{1948}{1}.

\bibitem{Ziegler-McNee-Konstan-Lausen-2005-WWW}
\Name{Ziegler C.~N., McNee S.~M., Konstan J.~A. \and Lausen G.}
  \Book{{Improving recommendation lists through topic diversification}} in
  \Book{Proceedings of the 14th International Conference WWW} (ACM, New York)
  2005 pp. 22--32.

\end{thebibliography}

\end{document}